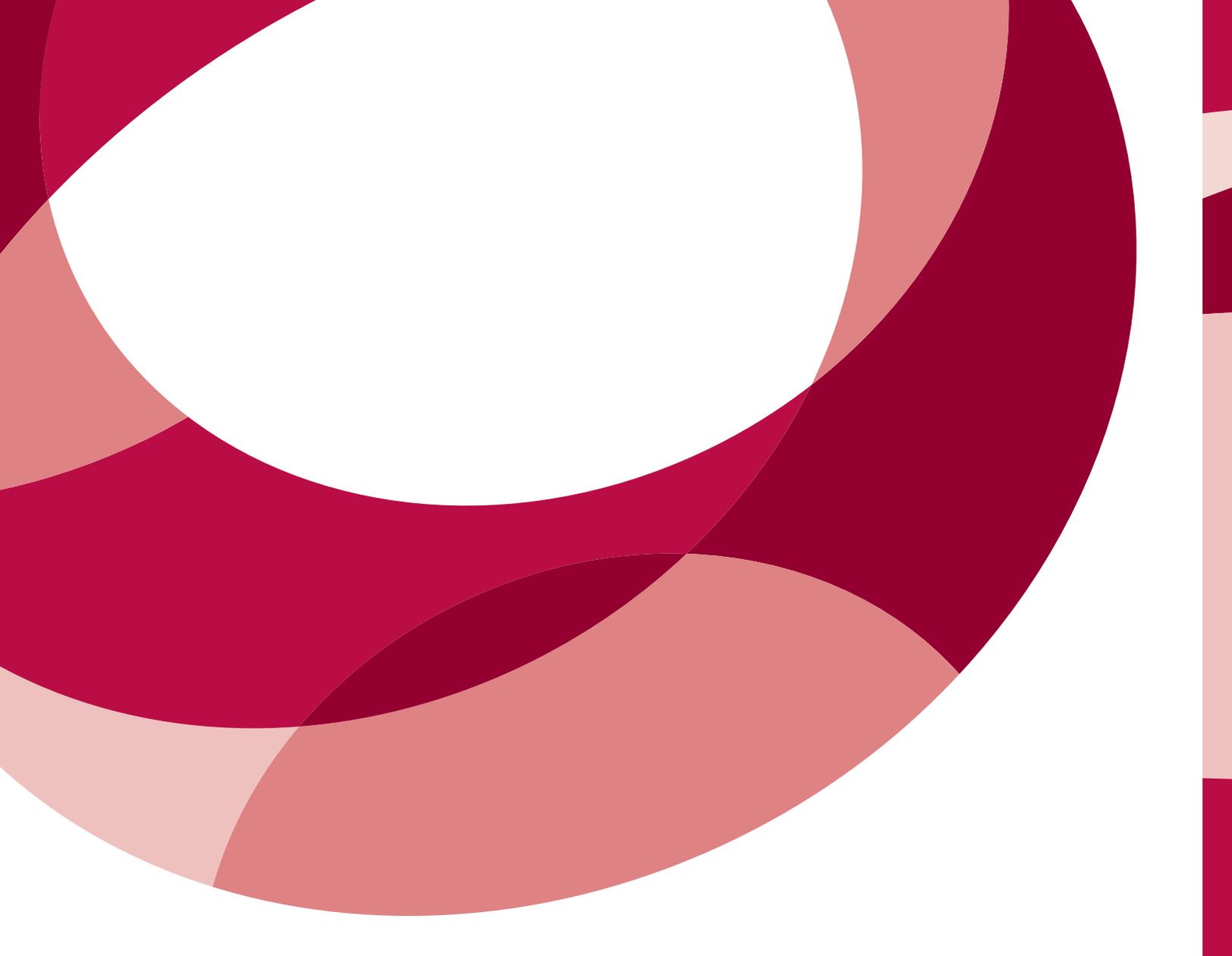

# Wide-Area Data Analytics

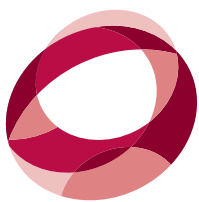

CCC
Computing Community Consortium
Catalyst

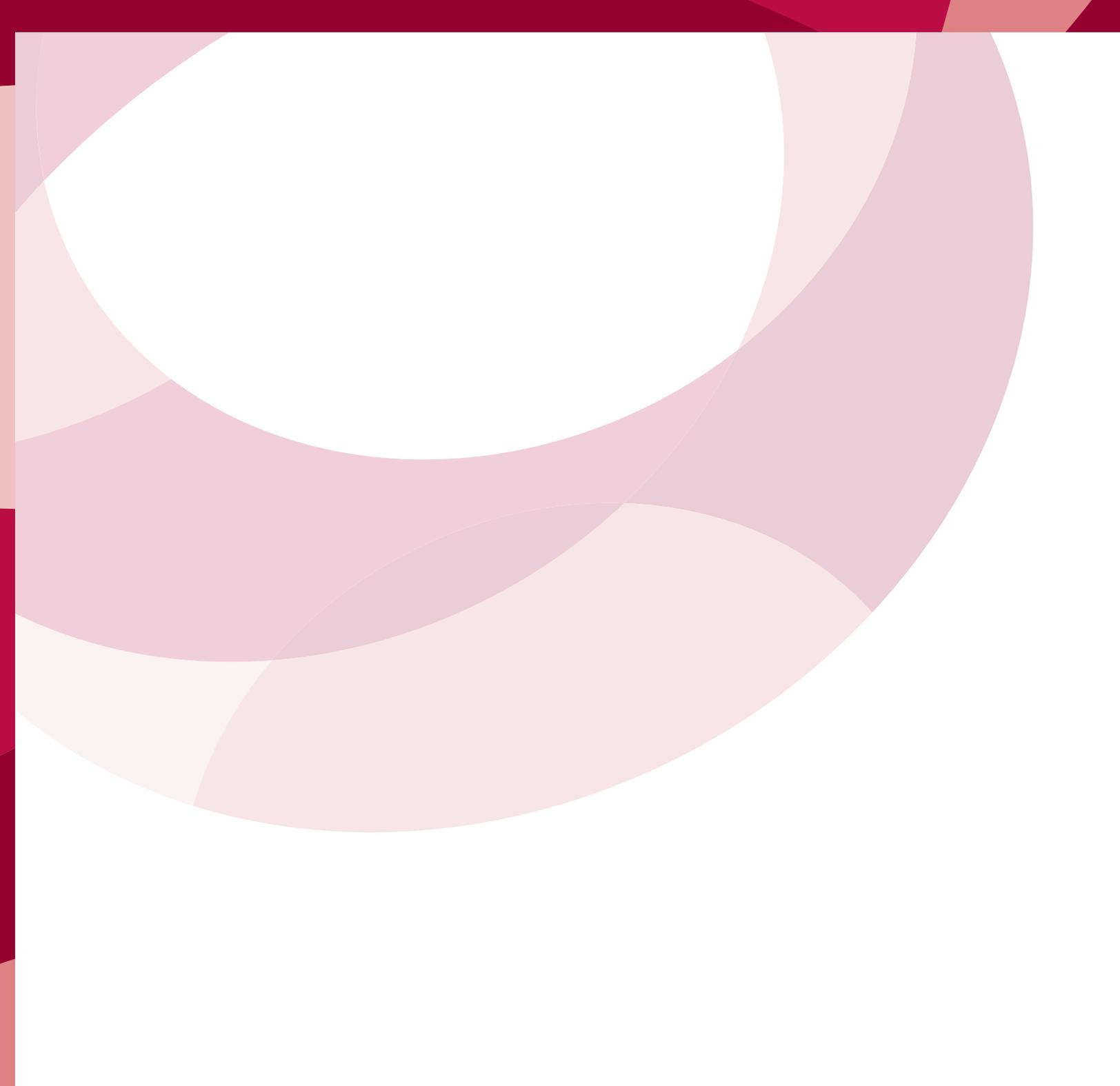

The material is based upon work supported by the National Science Foundation under Grants No. 1136993 and No. 1734706. Any opinions, findings, and conclusions or recommendations expressed in this material are those of the authors and do not necessarily reflect the views of the National Science Foundation.

# Wide-Area Data Analytics

A report based on a CCC workshop held October 3-4, 2019

Rachit Agarwal and Jennifer Rexford (workshop co-chairs) with contributions from numerous workshop attendees

June 2020

Sponsored by the Computing Community Consortium (CCC)

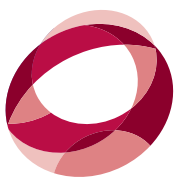

CCC
Computing Community Consortium
Catalyst

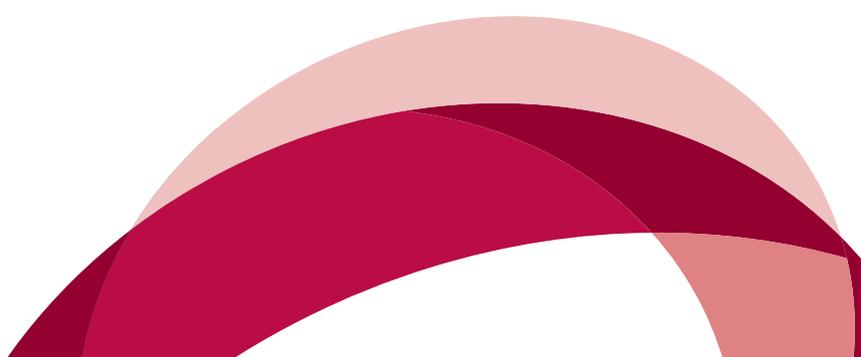

# 1. Introduction

We increasingly live in a data-driven world, with diverse kinds of data distributed across many locations. In some cases, the datasets are *collected* from multiple locations, such as sensors (e.g., mobile phones and street cameras) spread throughout a geographic region. The data may need to be analyzed close to where they are produced, particularly when the applications require *low latency* (e.g., augmented/virtual reality and self-driving cars), *high reliability* (e.g., oil rigs with no reliable communication channel, and self-driving cars that must provide basic functionality even when communication is disrupted), *low cost* (e.g., video feeds that are expensive to backhaul to a central location), *user privacy* (e.g., analyzing data on the user's mobile phone to avoid sending personal data to a cloud server), and *regulatory constraints* (e.g., sensitive data that must stay within a region with similar privacy frameworks like GDPR). In other cases, large datasets are *distributed* across public clouds, private clouds, or edge-cloud computing sites with more plentiful computation, storage, bandwidth, and energy resources. Often, some portion of the analysis may take place on the end-host or edge cloud (to respect user privacy and reduce the volume of data) while relying on remote clouds to complete the analysis (to leverage greater computation and storage resources).

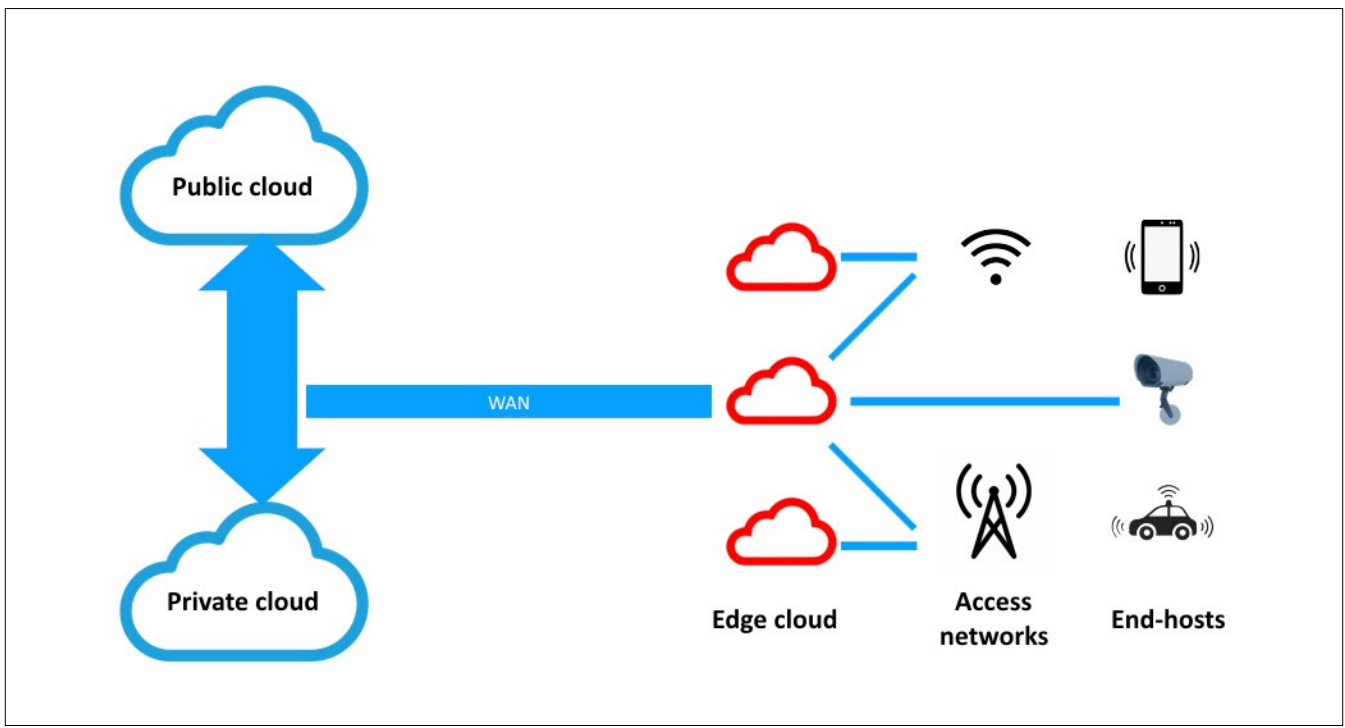

*Figure 1: Distribution of data across locations and devices.*

Wide-area data analytics is any analysis of data that is generated by, or stored at, geographically dispersed entities. Over the past few years, several parts of the computer science research community have started to explore effective ways to analyze data spread over multiple locations. In particular, several areas of "systems" research — including databases, distributed systems, computer networking, and security and privacy — have delved into these topics. These research subcommunities often focus on different aspects of the problem, consider different motivating applications and use cases, and design and evaluate their solutions differently. To address these challenges the Computing Community Consortium (CCC) convened a 1.5-day workshop focused on wide-area data analytics in October 2019. The workshop brought together researchers in these areas to identify a larger research agenda and uncover promising areas for collaboration. The workshop also included researchers in other core areas of computer science (e.g., machine learning, programming languages, computer architecture, and theory) that can inform the design of platforms for wide-area data analytics. This report summarizes the challenges discussed and the conclusions generated at the workshop.

The main theme that emerged from the workshop was the systems research challenges arising from *extreme heterogeneity*, including:

- End-host devices collecting the data (e.g., small sensors, video cameras, mobile phones, self-driving cars, and drones) and the resulting types of data (e.g., text, image, sensor readings, video streams),
- Computation, storage, and energy resources (e.g., small on end-hosts, modest in the edge cloud, and large in the data center),
- Access network technologies (e.g., wired broadband, WiFi, LTE, and 5G) and network bandwidth and latency (e.g., low in the access network and high across the wide area),
- Ownership and control of infrastructure by the users (e.g., end-hosts or home gateways), Internet Service Providers (e.g., access networks), enterprises (private cloud), or cloud providers (e.g., edge cloud or large data centers),
- Applications demanding low latency (e.g., augmented/virtual reality), high bandwidth (e.g., video analytics), and high reliability (e.g., self-driving cars), and
- Constraints on access to the data (e.g., to protect user privacy, to protect proprietary data, to obey regional regulations on data access and storage).

The workshop identified the following high-level "grand challenge": *Write a data-analysis question in a high-level language, without regard for where the data are collected, stored, or analyzed, and it "just works" — all while grappling with the extreme heterogeneity under the hood.* The results of the analysis should be *explainable* so the user understands the meaning of the results and the data used to compute them.

In the sections that follow, we first review several motivating use cases for wide-area data analytics. Then, we discuss research challenges and opportunities in several areas of "systems", including distributed systems, databases, networking, and security and privacy. Next, we outline several "cross-cutting" research areas, including languages for specifying policies for data access and storage, enabling a data-sharing marketplace, and supporting a common execution environment across heterogeneous end-host devices. We end by discussing ways to foster the research community through realistic benchmarks, academia-industry collaboration, and interdisciplinary research across the systems "stack."

## 2. Use Cases

The workshop attendees identified several use cases, including:

- Video analytics for real-time city surveillance, airport security, traffic monitoring, sporting events, and more;
- Analysis of diverse sensor data for monitoring wildlife, agriculture, laboratory experiments, manufacturing, offshore drilling, hospitals, and more;
- Augmented reality and virtual reality;
- Autonomous vehicles such as self-driving cars and drones;
- Sharing of sensitive data (e.g., personal fitness data, medical test results, financial data, information about security breaches) with analytics services, including combining data from multiple users or organizations to help create better models;
- Offloading of traditional data-analysis applications to the public cloud.

These use cases raise three interrelated challenges: latency constraints (for real-time applications), bandwidth constraints (for large datasets or devices with limited connectivity), and security and privacy constraints (imposed by users, companies, and governments).

## 3. Research Challenges and Opportunities

Wide-area data analytics introduces new opportunities and challenges across several areas of systems research, including distributed systems, databases, computer networking, and security and privacy.

### 3.1 Distributed Systems

Distributed systems form the "narrow waist" of modern wide-area services, offering simple and intuitive abstractions to applications while hiding the challenges of running over a distributed collection of heterogeneous devices. For example, applications may adopt familiar interfaces, such as Spark or Python, and implicitly assume that all of the data resides in one location with substantial computation and storage resources.

Under the hood, a distributed system manages data reads and writes — as well as computation, storage, and communication resources — across a range of devices, to hide latency, tolerate failures, and achieve scalability.

While these are not new challenges, emerging use cases make them even harder to address:

◗ The data sets are growing *larger*, due to the continuous collection of high-resolution sensor data (e.g., video) at many locations, as well as the joining of multiple kinds of data in a single analysis.

◗ The data are growing *more distributed*, including across devices in a single location (e.g., a data center), across multiple data centers, and at the points of collection (e.g., a self-driving car or a mobile phone).

◗ The distributed devices are *more heterogeneous*, including low-power sensors, mobile phones, programmable network elements, racks of high-end servers, and entire data centers.

◗ The network conditions are growing *more heterogeneous*, including disconnected operation (e.g., a self-driving car out of range of a cellular network), low-bandwidth access networks, and high-bandwidth core and data-center networks.

◗ The applications are growing more *latency-sensitive*, including interactive applications such as cyber-physical systems, or augmented and virtual reality. These applications may not tolerate the delay required to backhaul data to the cloud.

◗ The applications are growing more *heterogeneous*, with different tolerances for trading bandwidth and compute resources for accuracy.

◗ The applications impose more *constraints on data movement*, due to user privacy, proprietary data, national laws, and bandwidth limitations or costs.

Going forward, new research in distributed systems can explore a variety of exciting topics, including:

◗ *Characterizing emerging applications* — in terms of their performance requirements, acceptable costs, and tolerance for trading accuracy for better cost or performance — can drive the design of new interfaces and distributed systems to support them.

◗ *New abstractions and interfaces* can address new classes of applications and the changes in the underlying devices and network conditions. These interfaces can address the increasing heterogeneity of the devices, the need for low latency, the ability to trade accuracy for better cost or performance, and constraints on data movement.

◗ *Automating the partitioning of applications* across devices can simplify application development while still addressing the constraints on data movement, as well as optimizing for performance, reliability, cost, and energy-efficiency.

◗ *Exploiting emerging technologies* — such as domain-specific processors (e.g., GPUs, TPUs, and programmable packet processors) and non-traditional computing platforms (e.g., quantum computing and reconfigurable analog devices) — can help improve performance, cost, and energy-efficiency for future distributed systems.

### 3.2 Databases

Database systems allow users to efficiently interact with the data. Users write their queries using a standard language (in many cases, SQL), oblivious to how the data is stored and queried. Under the hood, the database system manages data storage, metadata (e.g., indexes) storage and management, and efficient query execution via highly optimized query plans while guaranteeing the semantics requested by the user (e.g., transactions, linearizability, etc.).

Designing databases for wide-area data analytics faces many of the same challenges affecting the design of distributed systems (mentioned above). Going forward, database research can explore a variety of exciting new topics:

◗ *Unified query language:* SQL has been the de-facto standard for querying structured and unstructured data. Do we need new languages for new kinds of data types, e.g., videos, images, etc. that we expect to be common in wide-area analytics? What is the right query language to look for events or objects in video? Do we need a natural language interface? If so, what are the basic set of data operations (filtering, join, spatial, time series) that need to be implemented? Can we extend existing SQL databases or Array/Spatial databases to allow users to declaratively define events in a video scene?

◗ *New algorithmic tools:* Large, heterogeneous, and diverse data raise new challenges for efficient indexing and computation. For some real-time applications, latency constraints may require analyzing data in a streaming fashion. For devices with limited memory or storage resources, the analysis may need to use compact data structures that reduce space requirements at the expense of some loss in accuracy.

◗ *Unified storage stack:* If data is to be combined from different sources (e.g., for building ML models that take video data from self-driving cars (LiDAR) as well as from street cameras) then the infrastructure needs to allow storage of and access to data from different sources. This presents related challenges of constructing a global or composed view of data for users and also for programmers to develop such mechanisms.

◗ *Query planning and optimization:* Providing mechanisms and an infrastructure for fast data analytics close to the edge remains a challenge. We need new mechanisms for distributing computation between the edge and the cloud that find the right tradeoff between bandwidth, computation, and latency (e.g., detecting public safety issues in real time from camera feeds).

## 3.3 Networking

Computer networks enable fast, reliable, and secure communication between users and their wide-area services, and among the many distributed components that work together to provide these services. For example, the suite of Internet protocols enable best-effort packet delivery (e.g., Ethernet and WiFi in the local area, and IP across the wide area), end-to-end connections (e.g., byte stream and datagram delivery), and the exchange of application messages (e.g., the web, e-mail, telephony, domain name lookups, and more). These protocols rely on interoperable implementations across a wide array of devices, as well as high-speed mechanisms for processing, buffering, and forwarding data packets and a range of physical media offering good performance and reliability in transferring data.

New and emerging applications make the design and operation of computer networks more challenging:

◗ The sources of data are *increasingly mobile,* including smart phones, self-driving cars, drones, and sensors on animals, products, and equipment. Yet, mobile devices tend to have poor and widely varying performance, due to frequent changes in network conditions and the difficulties of maintaining end-to-end communication on the go.

◗ New applications, such as cyberphysical systems and AR/VR, require *low latency* communication. Yet, network protocols often require multiple round-trip times to establish connections, bootstrap security, and retransmit lost data.

◗ End-host and network devices, and the physical media along the paths between them, are increasingly *heterogeneous* in cost, performance, reliability, and energy-efficiency. This makes it difficult to design protocols that work well — and can be implemented efficiently — across a range of components and settings.

◗ Modern applications impose increasingly strong demands for *privacy and security,* leading to more requirements on network protocols to sign and encrypt data, protect user anonymity, prevent surveillance, and detect and block cyberattacks.

◗ The increase in networked devices, particularly from mobile devices and sensors, introduces *scalability* challenges. Network devices and protocols must handle communication for a larger number of devices, with more frequent mobility, transferring larger volumes of data, over higher-speed links.

Going forward, new research in computer networking can explore a variety of exciting topics:

◗ The emergence of *5G access networks* offers the promise of higher bandwidth and lower latency, as well as network "slicing" to isolate and customize networking for different tenants. These access networks will often have local computation and storage resources, enabling co-design of network protocols and the distributed services that run over them.

◗ To support mobile devices and heterogeneous network conditions, network protocols should support ways for devices to *process data locally until better (or cheaper) connectivity is available.* New research can explore techniques for predicting future connectivity, and the design of network architectures with sporadic connectivity in mind.

◗ Emerging network interface cards and switches offer much *greater programmability,* offering greater flexibility without sacrificing speed. These devices can support network functionality normally relegated to slower software components (such as middleboxes or control-plane protocols) or distributed systems, and also offer better visibility into network conditions such as performance problems or cyberattacks.

◗ Future network designs can *refactor the "division of labor"* between end-hosts and the network devices. For example, end-host mobility could be handled more efficiently by the end-host network stack (e.g., connection migration mechanisms in the QUIC transport protocol) rather than by the underlying network.

◗ Internet providers offer *service-level agreements* (SLAs) for network properties such as throughput, loss, and latency over large timescales. However, SLAs for data-analytics applications could look quite different, focusing on latency on smaller timescales as well as metrics related to the quality of the data-analytics results.

◗ *The Internet is growing more flat*, with cloud services, content distribution networks, and edge servers increasing close to the end users. In addition, a smaller set of organizations offer and manage these services. The flattening of the network could simplify the deployment of new networking technologies, while raising new concerns about security, privacy, and robustness.

## 3.4 Security and Privacy

Security is the protection of computer, storage, and networked systems from damage or theft, as well as from the disruption or misdirection of the services they offer. Security is often discussed in the context of the CIA triad of *confidentiality* (ensuring information is not disclosed to unauthorized individuals), *integrity* (ensuring the accuracy and completeness of data), and *availability* (preventing service disruptions due to outages, failures, and attacks). In contrast, privacy is the protection of individuals and organizations from having their data or other information (e.g., personally identifiable information, trade secrets, etc.) disclosed to others without permission. Both security and privacy have grown significantly in importance in recent years, due to high-profile attacks by individuals and nation-states, as well as concerns about how companies collect and analyze user data. Wide-area data analysis raises a number of additional privacy and security challenges:

◗ Geo-distributed applications increasingly collect *sensitive or personal data* to provide various kinds of services (e.g., Uber, Lyft, social networks, street cameras).

◗ Advances in data-analysis techniques (e.g., face recognition) make it increasingly easy to *infer sensitive information* from raw data.

◗ New applications and the popular press are *raising user awareness* about privacy and security concerns, and the economic value of user data.

◗ *Privacy policies* like GDPR (General Data Protection Regulation) are becoming more prevalent, and can differ across countries and regions. The emergence of these policies also drive user awareness and concern about privacy.

◗ Geo-distributed applications may analyze or combine data from *multiple owners*, who each want *control* over how their data are used.

◗ The underlying distributed compute, storage, and network infrastructure need good *security properties* — individually and in how they work in concert — to protect user data and ensure that critical data-driven applications operate correctly and reliably.

Allowing users and organizations to express control, and creating both a *policy* and a *systems* infrastructure to enforce it, is the main problem posed in terms of secure, private, wide-area data analytics. This requires advances in:

◗ *New languages* for users and organizations to express policies on how their data can be used, potentially in a dynamic manner, as consent to use data may change over time (e.g., the right to be forgotten). For example, some people may want to contribute their medical records for cancer research but not for tobacco research; or some people may allow their images to be used to train neural networks for facial recognition but not made available more broadly.

◗ *New security mechanisms* are needed to enforce these policies. The guarantees provided should be system-wide guarantees (e.g., we need guarantees not just on raw data, but also on derived data, and across the collection of distributed computational, storage, and bandwidth resources).

◗ *New tools* are necessary to track data movement across the components of the system, and to detect policy violations. Upon detecting violations, these tools should automatically bring the system to the "right" state (e.g., retroactively deleting the corresponding data as well as derived data and metadata).

◗ *A federated data marketplace* could offload computation to where data is located. For instance, in the case of a rare disease where each hospital has a local, limited, view of the disease, a marketplace may allow models to be trained at each individual hospital (thus maintaining privacy), and yet combined together to produce a global model.

◗ Techniques for generating and publishing *synthetic data* sets based on generative models trained on the original data (e.g., GANs) can help foster better research without divulging sensitive user information.

◗ Cryptographic tools like *Secure Multi-Party Computation* (SMPC) can enable distributed analytics for sensitive data. Several open issues here include the scalability of SMPC (e.g., for complex learning tasks) and privacy guarantees provided by combinations of techniques (e.g., SMPC with differential privacy).

## 4. Cross-cutting Systems Research Challenges

In addition to systems challenges, wide-area data analytics raises a number of cross-cutting research issues.

### 4.1 Policy and Systems Infrastructure

A policy infrastructure will impose requirements on the systems infrastructure, but the capabilities of the system infrastructure will also influence policy. We expect the two to co-evolve. Much of the existing policy, e.g., GDPR, are aspirational (declarative) in the sense that they describe a high-level goal without regard for the underlying technology required to achieve it. Some of this technology is not present today. For example, GDPR asserts the right to be forgotten and the right to data portability, but the systems infrastructure required to support both is currently not available (at least not fully). GDPR also distinguishes between (raw) personal data and derived data (e.g., statistics or machine-learning models generated from their data), and does not provide much protection over derived data, but it is unclear if this is consistent with what users want or how difficult it would be for systems infrastructure to provide.

### 4.2 Data Sharing and Market

Data collection has become increasingly pervasive and heterogeneous over the last few years. However, today the usefulness of the collected data is limited for two reasons: (1) data typically resides in administrative silos within organizations, across organizations, on personal devices, etc.; and (2) the infrastructure necessary to facilitate the exchange, sharing, and processing of data is largely non-existent. Huge societal, scientific, and business benefits can be reaped by the combination of data from multiple sources and of multiple types. For example, rare disease models that exploit data from multiple hospitals/genomics labs could be much more powerful than those that are restricted to a single hospital/lab. In general, combining data from multiple sources could result in value that can be much larger than the sum of the parts; moreover, the cost of data acquisition can/should be amortized across multiple use cases, users, and applications. To realize this vision, we will need interdisciplinary advances:

◗ We need platforms that can efficiently support a data and services ecosystem of data owners (organizations or individuals), data consumers (e.g., research organizations that want to train models on the data), and data processors (third-party organizations offering services on the data, e.g., anonymization, harmonization, federated learning). Such a platform would require a number of components, including, policies, systems and query languages for: (i) data and service discovery, (ii) data integration/federation (e.g., different data sources may have different formats or schemas), (iii) identifying data provenance, quality, timeliness, and integrity, (iv) data tracking, usage monitoring, and auditing, and (v) enforcing the chain of custody.

◗ We need new mechanisms that incentivize data sharing (e.g., new pricing mechanisms) from organizations and individuals. The incentives for organizations may be very different than those for individuals.

◗ Data sharing and market platforms must guarantee privacy and security. The platform must incorporate mechanisms for tracking and enforcing user consent, both within the raw data as well as derived data. To that end, such a platform would greatly benefit from advances in usable and scalable privacy and security mechanisms, as well as systems and languages for privacy-preserving data analytics.

## 4.3 A Common Execution Environment for Heterogeneous Devices

Wide-area data analytics is made more complicated by the tremendous heterogeneity of devices, including sensors, IoT devices, drones, mobile phones, network interface cards and switches, high-end servers with hardware accelerators, and more. In addition to having different computational, bandwidth, and storage resources, these devices also have different operating systems and software platforms, making it very difficult to create and deploy analytics software. Creating a common execution environment across a range of devices would enable a "write once, deploy everywhere" principle. The core question is whether the common execution environment should be an API, a container, or a full virtual machine.

Answering this question may be gated upon how the field and the definition of "edge" evolves: for a definition that includes all devices all the way to the users, the common execution environment might need to be lean and light-weight; for a definition that excludes the last mile (e.g., excludes user devices but includes the last-hop router or access point), the common execution environment could be more powerful since these devices are relatively less resource constrained; finally, for a definition that includes the cloud provider to operate all infrastructure leading up to the edge, the constraints may be more pragmatic. Nevertheless, the systems and the database communities should explore the feasibility of such a common execution environment.

## 5. Fostering the Research Community

Wide-area analytics is likely to lead to a diverse application community that is trying to test applications on diverse hardware and infrastructure. It is hard, if not impractical, to accommodate experiments and evaluations for all applications within a unified framework and infrastructure. We need new "stacks" for assembling benchmarks and designs, and for implementing testbeds and platforms that can fuel research in this emerging area. At this early stage, it is also important to give particular consideration to reproducible research so as to enhance the usability and portability, especially enabling the platform/testbed to adapt to the development of hardware. To that end, we need a community-wide effort to:

◗ Create benchmarks, testbeds, and open-source platforms that allow researchers to build on each other's work and make solutions broadly available.

◗ Create relationships between academia and industry, industry consortia, etc. to align academic progress with industrial deployments; and, to encourage open ecosystems in an otherwise increasingly diverse and heterogeneous industrial deployments.

◗ Encourage more interdisciplinary efforts and funding to fuel university-led end-to-end research, from design to implementation to deployments.

## 6. Conclusions

Advances across the entire "systems stack" are needed to address the grand challenge of wide-area data analytics: *Enabling users to write data-analysis questions in a high-level language, without regard for where the data are collected or stored, and everything "just works".* We believe that research that cuts across the traditional "layers" of the stack, and is driven by compelling use cases and the capabilities of emerging devices, can lead to unprecedented progress toward this ambitious goal.

# 7. Appendix
## Workshop Participants

| First Name | Last Name | Affiliation |
|---|---|---|
| Rachit | Agarwal | Cornell University |
| Ganesh | Ananthanarayanan | Microsoft Research |
| Mina | Arashloo | Princeton University |
| Victor | Bahl | Microsoft Research |
| Sujata | Banerjee | VMware |
| Xiaoqi | Chen | Princeton University |
| Mosharaf | Chowdhury | University of Michigan |
| Khuzaima | Daudjee | University of Waterloo |
| Khari | Douglas | CRA/CCC |
| Michael | Franklin | University of Chicago |
| Mike | Freedman | Princeton University |
| Minos | Garofalakis | Technical University of Crete |
| Qiang | Guan | Kent State University |
| Bo | Han | AT&T |
| Mark | Hill | University of Wisconsin-Madison |
| Wenjun | Hu | Yale University |
| Xin | Jin | Johns Hopkins University |
| Anurag | Khandelwal | Yale University |
| Boon Thau | Loo | University of Pennsylvania |
| Deepankar | Medhi | National Science Foundation |
| Krzysztof | Onak | IBM Research |
| Manish | Parashar | National Science Foundation |
| Dan | Ports | Microsoft Research |
| Lili | Qiu | University of Texas at Austin |
| Jennifer | Rexford | Princeton University |
| Stefan | Robila | National Science Foundation |
| Mohamed | Sarwat | Arizona State University |
| Ann | Schwartz Drobnis | CRA/CCC |
| Siddhartha | Sen | Microsoft Research |
| Jonathan | Smith | DARPA |
| Ion | Stoica | University of California, Berkeley |
| Midhul | Vuppalapati | Cornell University |
| Helen | Wright | CRA/CCC |
| Minlan | Yu | Harvard University |

# NOTES

**NOTES**

**NOTES**

___________________________________________________

___________________________________________________

___________________________________________________

___________________________________________________

___________________________________________________

___________________________________________________

___________________________________________________

___________________________________________________

___________________________________________________

___________________________________________________

___________________________________________________

___________________________________________________

___________________________________________________

___________________________________________________

___________________________________________________

___________________________________________________

___________________________________________________

___________________________________________________

___________________________________________________

___________________________________________________

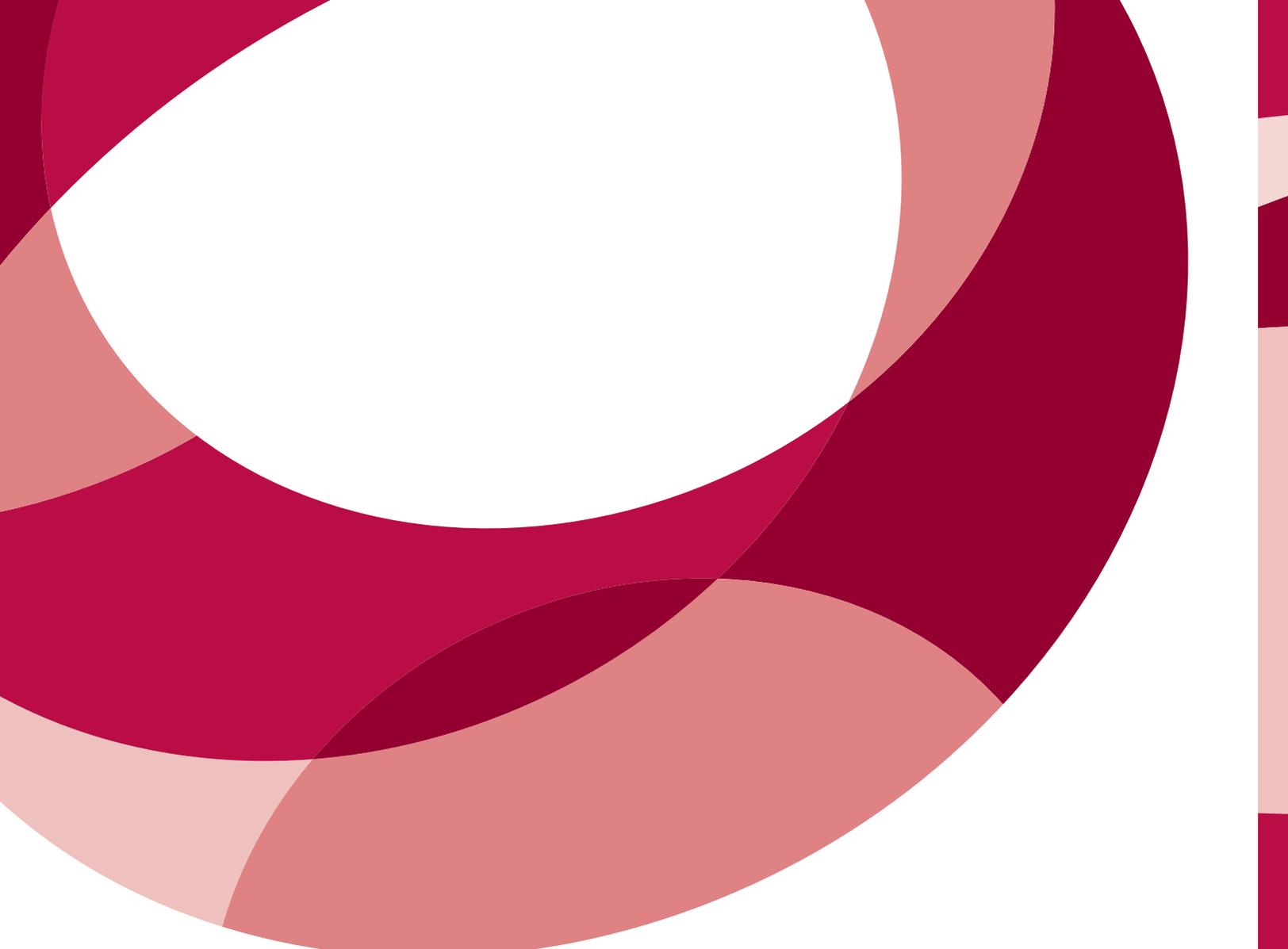
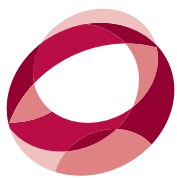
**CCC**
Computing Community Consortium
Catalyst

1828 L Street, NW, Suite 800
Washington, DC 20036
P: 202 234 2111 F: 202 667 1066
www.cra.org cccinfo@cra.org